\documentclass[aps,prl,reprint]{revtex4-1}  
\usepackage{graphicx}  
\usepackage{dcolumn}   
\usepackage{multirow}
\usepackage{bm}        
\usepackage{amssymb}   
\usepackage{amsmath}   
\usepackage{siunitx}   
\usepackage{color}     

\usepackage{blindtext}

\usepackage[colorinlistoftodos]{todonotes}
\usepackage{verbatim}
\usepackage[normalem]{ulem} 

\DeclareSIUnit\torr{Torr}
\DeclareSIUnit\sq{\ensuremath{\Box}}

\begin{document}


\title{High Kinetic Inductance NbN Nanowire Superinductors}

\author{David~Niepce}
\email{david.niepce@chalmers.se}
\author{Jonathan~Burnett}
\author{Jonas~Bylander}
\affiliation{Chalmers University of Technology, Microtechnology and Nanoscience, SE-41296, Gothenburg, Sweden}
\vskip 0.25cm

\date{\today}

\begin{abstract}
We demonstrate that a high kinetic inductance disordered superconductor can realize a low microwave loss, non-dissipative circuit element with an impedance greater than the quantum resistance ($R_Q = h/4e^2 \simeq \SI{6.5}{\kilo\ohm}$). This element, known as a superinductor, can produce a quantum circuit where charge fluctuations are suppressed. 
The superinductor consists of a $\SI{40}{\nano\meter}$ wide niobium nitride nanowire and exhibits a single photon quality factor of $\SI{2.5e4}{}$. Furthermore, by examining loss rates, we demonstrate that the dissipation of our nanowire devices can be fully understood in the framework of two-level system loss.
\end{abstract}

\maketitle

Disorder within superconductors can reveal non-trivial electrodynamics~\cite{Driessen2012,MondalNbN}, dual Josephson effects~\cite{astafiev2012}, and superconducting-insulating phase transitions (SIT)~\cite{HavilandSIT}. In general, the disorder increases the superconductors' normal-state resistance, which also enhances the kinetic inductance. High kinetic inductance can be used to produce circuits with impedances exceeding the resistance quantum ($R_Q\,=\,h/4e^2\,\simeq\,\SI{6.5}{\kilo\ohm}$). A quantum circuit element with zero dc resistance, low microwave losses and an impedance above $R_Q$ is known as a superinductor~\cite{masluk2012,bell2012}. Qubits based on superinductors are immune to charge fluctuations and have demonstrated extraordinary relaxation times~\cite{pop2014}. However, these examples of superinductors have been based on the kinetic inductance of Josephson junction arrays (JJA), which places constraints on the possible device parameters and geometries. Therefore, there remains interest in a superinductor formed by a disordered superconducting nanowire~\cite{kerman2010}. 

A disordered superconductor-based superinductor should possess tremendous magnetic field tolerance~\cite{samkharadze2016}, suitable for hybrid qubits which operate at high magnetic fields~\cite{luthi2017evolution}. To fully exploit these circuits, more work is needed to understand and mitigate sources of unconventional dissipation relating to the proximity to the SIT~\cite{coumou2013microwave,feigel2017microwave}. This motivates the need to fabricate a sufficiently disordered superconductor to obtain high inductance, but not so disordered as to induce dissipation. Therefore, to maximize impedance, it becomes crucial to minimize the stray capacitance, which can be achieved with a nanowire geometry. However, while low-loss superconducting nanowires can be fabricated~\cite{burnett2017}, the nanowire geometry is itself more susceptible to parasitic two-level systems (TLS). This susceptibility is due to an enhanced participation ratio~\cite{gao2008} of the TLS host volume to the total volume threaded by the electric field. It is due to these loss mechanisms that present superinductor circuits have preferred JJAs~\cite{masluk2012} rather than disordered superconductors. 

Here, we demonstrate a nanowire-based superinductor with an impedance of $\SI{6.795}{\kilo\ohm}$. We developed a process~\cite{Note2} based on dry etching a hydrogen silsesquioxane (HSQ) mask to pattern a $\SI{20}{\nano\meter}$ thick, strongly disordered film of niobium nitride (NbN) into a $\SI{40}{\nano\meter}$ wide and $\sim\!\SI{680}{\micro\meter}$ long nanowire.
These dimensions ensure a large inductance while exponentially suppressing unwanted phase slips in our devices~\cite{Note2,peltonen2013}. We study both the microwave transmission and dc transport properties of several nanowires to characterize their impedance and microwave losses. These nanowire-based superinductors demonstrate a single photon quality factor of $\SI{2.5e4}{}$, which is comparable to JJA-based superinductors~\cite{masluk2012}. 
We find that the dominant loss mechanism is parasitic two level systems (TLS), which is exacerbated by the unfavorable TLS filling factor~\cite{gao2008} that arises from the small dimensions required to obtain a high impedance. 

\begin{figure}
    \includegraphics[width=8.5cm]{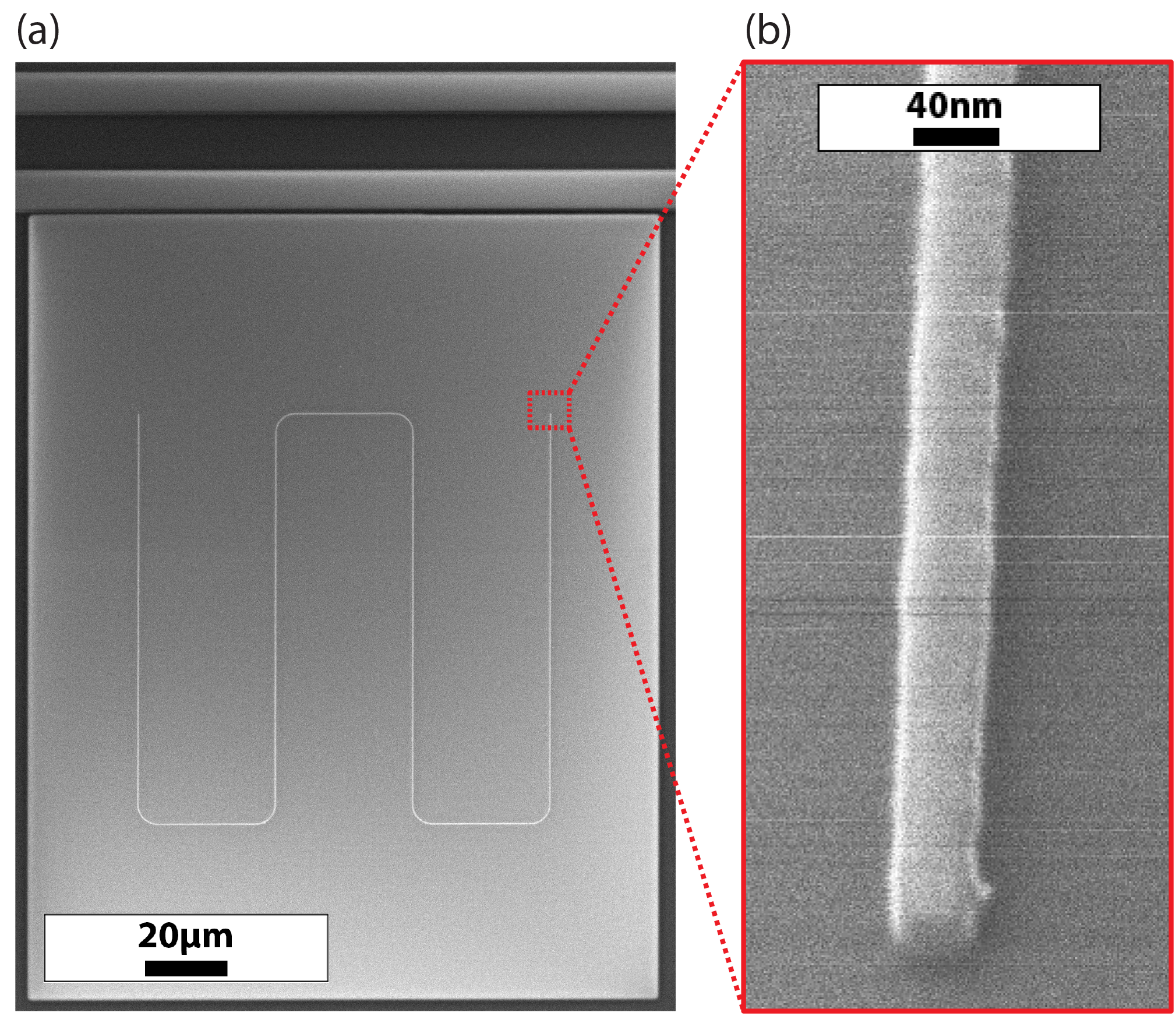}
    \includegraphics[width=8.5cm]{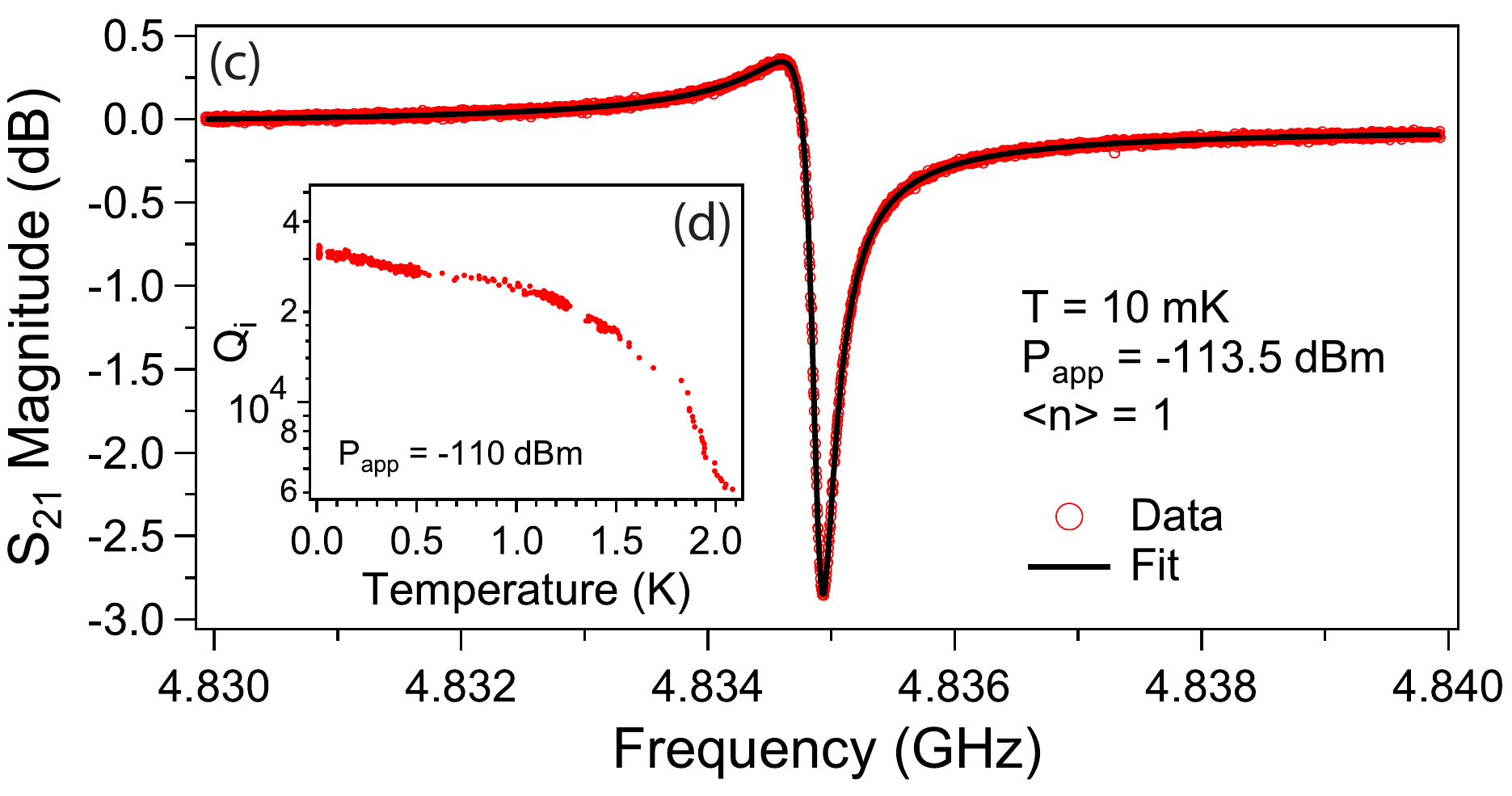}
    \caption{\label{fig:device}{\bf (a)} SEM micrograph of a nanowire resonator coupled to its feed line; the NbN feed line and ground plane are shown in black, the Si substrate is in gray, and the $\SI{40}{\nano\meter}\times\SI{680}{\micro\meter}$ nanowire is light gray.  
    {\bf (b)} A helium FIB image of the nanowire. 
    {\bf (c)} $S_{21}$ magnitude response of the device, in the single-photon regime. The black line is a fit to determine the resonance parameters. {\bf (d)} A plot of the internal quality factor $Q_i$ of a nanowire superinductor resonator as a function of temperature.}
\end{figure}

The final sample contains five nanowires that are inductively coupled to a common microwave transmission line. The sample also contains separate dc transport test structures. Figure~\ref{fig:device}(a--b) show micrographs of a typical device.
We study the microwave properties of these resonators by measuring the forward transmission ($S_{21}$) response.
Figure~\ref{fig:device}(c) shows a typical $S_{21}$ magnitude response measured at $\SI{10}{\milli\kelvin}$ and with an average photon population $\left< n \right> = 1$. 
We determine the resonator parameters by fitting the data with a traceable fit routine~\cite{probst2015}. From this we find a resonant frequency $f_r = \SI{4.835}{\giga\hertz}$ and an internal quality factor $Q_i = \SI{2.5e4}{}$. 

\begin{figure}[b]
    \includegraphics[width=8cm]{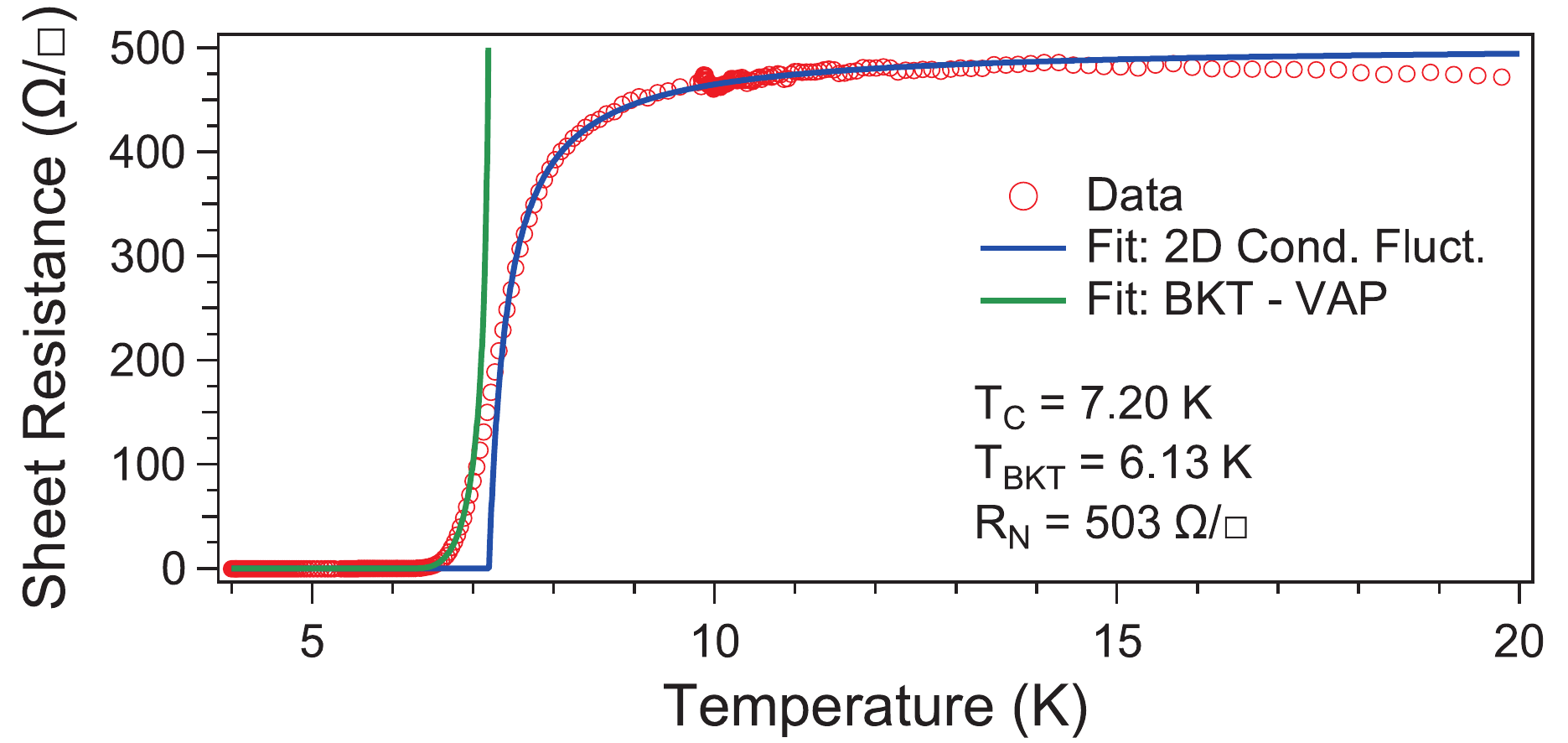}
    \caption{\label{fig:transport}$R(T)$ characteristic of an NbN nanowire. The blue and green lines are fits to Eqs.~(\ref{eq:AL})~and~(\ref{eq:BKT}) respectively.}
\end{figure}

To understand this resonance, we first study the transport properties of our NbN nanowires.
We estimate the kinetic inductance contribution $L_k^{\Box}(0)$ using~\cite{tinkham}
\begin{equation}
\label{eq:Lk}
L_k^{\Box}(0) = \dfrac{\hbar R_N^{\Box}}{\pi \Delta_0}
\end{equation}
\noindent where $R_N^{\Box}$ is the normal-state sheet resistance, $\hbar$ the reduced Planck constant, and $\Delta_0$ the superconducting gap at zero temperature. NbN is experimentally found to be a strongly coupled superconductor with $\Delta_0 = 2.08 k_B T_c$~\cite{MondalNbN}, where $T_c$ is the critical temperature and $k_B$ is the Boltzmann constant. Figure~\ref{fig:transport} shows a typical $R(T)$ characteristic of an NbN nanowire. 
From room temperature, as the temperature decreases, the resistance increases until a plateau is reached at about $\SI{15}{\kelvin}$. This behavior is typical of weak localization in strongly disordered materials~\cite{bergmann1983}. As the temperature further decreases from $\SI{15}{\kelvin}$, the resistance starts to decrease and we observe a $\SI{2}{\kelvin}$ wide superconducting transition.

The width of this superconducting transition can be fully described by two different mechanisms. Above $T_c$, thermodynamic fluctuations give rise to short-lived Cooper pairs, which increase the conductivity. These conductivity fluctuations have been described in the 2D case by Aslamasov and Larkin~\cite{tinkham} and are given by
\begin{equation}
\label{eq:AL}
\sigma_{2D}(T) = \dfrac{e^2}{16 \hbar d}\left( \dfrac{T_c}{T-T_c} \right)
\end{equation}
\noindent where $T$ is the temperature, $e$ is the electron charge and $d$ is the film thickness. The total conductivity above $T_c$ is now expressed as $\sigma(T) = \sigma_n + \sigma_{2D}(T)$. 

Below $T_c$, the resistance does not immediately vanish. This can be explained by a Berezinskii--Kosterlitz--Thouless (BKT) transition~\cite{mooij} where thermal fluctuations excite pairs of vortices. These vortex-antivortex pairs (VAP) are bound states, formed by vortices with supercurrents circulating in opposite directions. Above the ordering temperature $T_{BKT}$, VAPs start to dissociate and their movement cause the observed finite resistance. This resistivity is described by~\cite{mooij}
\begin{equation}
\label{eq:BKT}
\rho(T) = a \exp \left( -2 \sqrt{b\dfrac{T_c - T}{T - T_{BKT}}} \right)
\end{equation}
\noindent with $T_{BKT} < T < T_c$, and where $a, b$ are material dependent parameters. 

Fitting the $R(T)$ of Figure~\ref{fig:transport} to Eqs.~(\ref{eq:AL}--\ref{eq:BKT}) leads to $R_N^{\Box} = \SI[per-mode=symbol]{503}{\ohm\per\sq}$ and $T_c = \SI{7.20}{\kelvin}$. Using Eq.~(\ref{eq:Lk}), this yields $L_k^{\Box}(0) = \SI[per-mode=symbol]{82}{\pico\henry\per\sq}$. 
For a $\SI{40}{\nano\meter}$ wide nanowire this corresponds to an inductance per unit length of $\SI{2.05}{\milli\henry\per\meter}$. From an empirical formula~\cite{mohan1999}, the magnetic inductance, due to the geometry of the nanowire, is estimated to be only $L_{m} \simeq \SI{1}{\micro\henry\per\meter}$. Therefore, we assume the nanowire inductance arises entirely from the kinetic inductance, so that $L_{nw} = L_k$. Using Sonnet \textit{em} microwave simulator, we estimate the capacitance per unit length of our nanowires to be $C_{nw} = \SI{44.4}{\pico\farad\per\meter}$. Combining these properties leads to an estimated resonance frequency within 1\% of the measured resonance frequencies of our resonators. Using these parameters, we calculate the characteristic impedance of our nanowires to be $Z_r = \sqrt{L_{nw}/C_{nw}} = \SI{6.795}{\kilo\ohm} \pm \SI{35}{\ohm}$. Therefore, $Z_r \geq R_Q$, indicating that our nanowires are superinductors.

Having demonstrated superinductors, we now examine their behavior as a function of applied microwave drive and varying temperature. 
We first determine the range of temperatures at which we can operate our device. 
Figure~\ref{fig:device}(d) shows a measurement of the internal quality factor $Q_i$ against temperature. We show that from $\SI{10}{\milli\kelvin}$ to $\SI{1.4}{\kelvin}$, the quality factor only marginally decreases from $\SI{3e4}{}$ to $\SI{2e4}{}$. This offers a far greater range of operation than aluminum JJA-based superinductors, which show significant dissipation above 100~mK\cite{masluk2012}. 

We now investigate the low-temperature loss mechanisms as a function of microwave power. When probed with an applied power $P_{in}$, the average energy stored in a resonator of characteristic impedance $Z_r$ is given by $\left< E_{int} \right> = Z_0 Q_L^2 P_{in} / \pi^2 Z_r Q_c f_r$, where $Z_0 = \SI{50}{\ohm}$, $Q_c$ and $Q_L$ are respectively the coupling and loaded quality factors of the resonator. In the following, we describe the microwave power in $\left< n \right>$, the average number of photons in the resonator, given by $\left< n \right> = \left< E_{int} \right>/h f_r$ where $h$ is the Planck constant.

\begin{figure}
    \includegraphics[width=8.5cm]{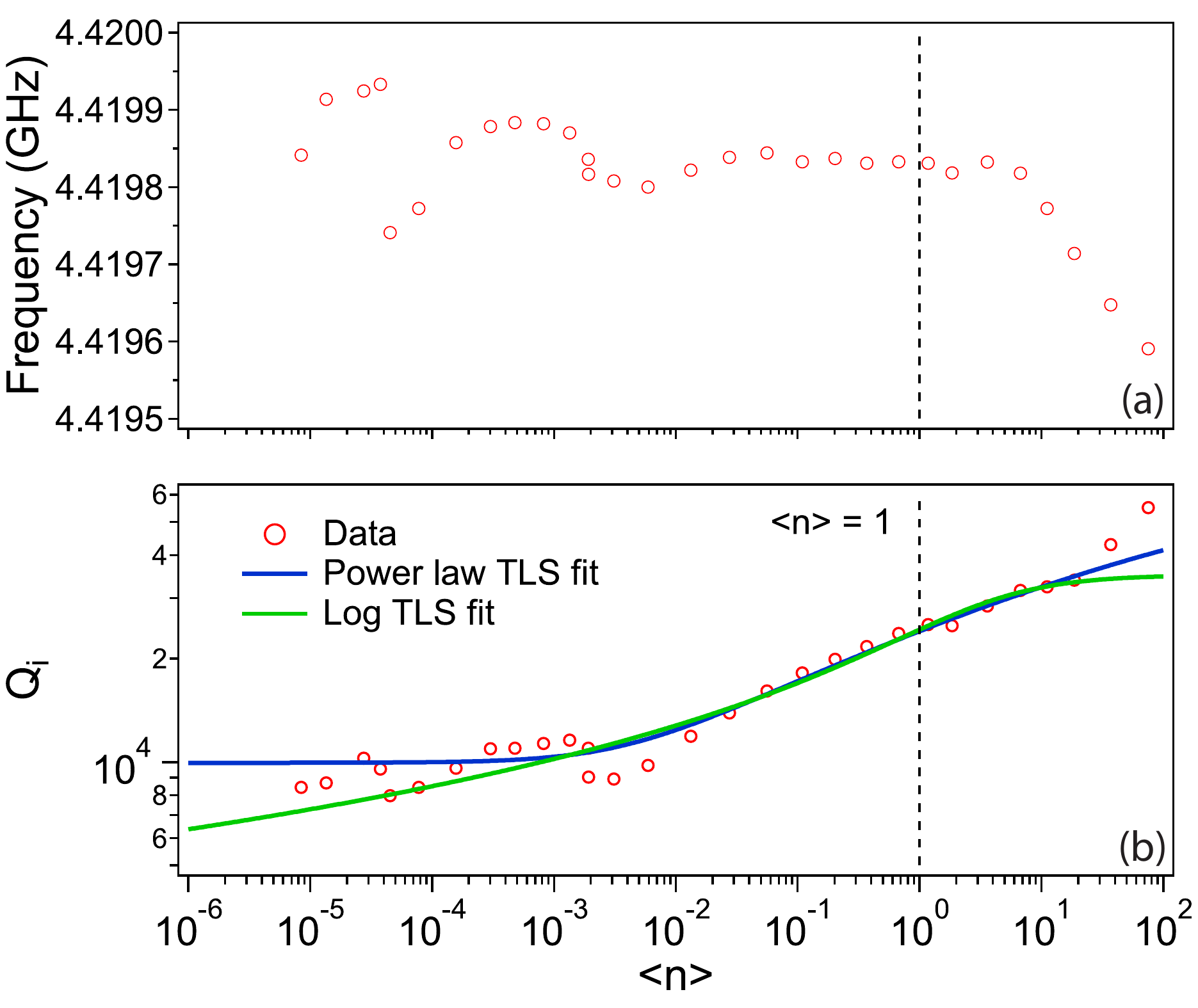}
    \caption{\label{fig:sweep}{\bf (a)} Resonant frequency of a typical nanowire resonator as a function of microwave drive.
    {\bf (b)} Internal quality factor ($Q_i$) of a typical nanowire superinductor resonator as a function of the microwave drive. Solid lines are fits to Eq.~(\ref{eq:TLSpow}) (blue) and Eq.~(\ref{eq:TLSlog}) (green).
    The vertical dashed lines highlight the single microwave photon regime.}
\end{figure}

Due to the large impedance of our resonator, we are able to measure in the low photon regime with a high applied power. Consequently, this enables us to measure, with good signal-to-noise ratio, photon populations two to three orders of magnitude lower than in conventional resonators. In Figure~\ref{fig:sweep}(a), starting at $\left< n \right> = 1$, as we increase power, the resonance frequency does not change until $\left< n \right> \simeq 10$. From $\left< n \right> \simeq 10$, as power increases, the frequency decreases until the resonator bifurcates. This is explained by the power dependence of the kinetic inductance, which behaves as a Duffing-like non-linearity~\cite{swenson}. We note that this non-linearity occurs at similar microwave drives as junctions-embedded resonators~\cite{osborn2007}. Starting again at $\left< n \right> = 1$, as we decrease power, we see the frequency remain approximately constant. Although, as $\left< n \right>$ is decreased below  $\left< n \right> \simeq 10^{-3}$, the resonator exhibits frequency jitter, consistent with TLS-induced permittivity changes\cite{burnett2014}. This frequency noise results in spectral broadening of the resonance curve.

We now examine the internal quality factor as a function of applied microwave power (shown in Figure~\ref{fig:sweep}(b)). Between the range of $\left< n \right> \simeq 10^{-5}$ and $\left< n \right> \simeq 10^{-3}$, we find that $Q_i$ is approximately constant, with changes in $Q_i$ caused by frequency jitter-induced spectral broadening. From $\left< n \right> \simeq 10^{-3}$, as we increase power, $Q_i$ increases, which is consistent with depolarization of TLS. For $\left< n \right> \geq 40$, $Q_i$ is overestimated due to the Duffing non-linearity. We fit our data to a common TLS loss model~\cite{burnett2017} described by
\begin{equation}
\label{eq:TLSpow}
\dfrac{1}{Q_i} = \delta^i_{tot} = F\delta^0_{TLS}\dfrac{\tanh\left(h f_r /2 k_B T\right)}{\left(1+\left<n\right>/n_c\right)^\beta} + \delta_{0}
\end{equation}
\noindent where $n_c$ is the number of photons equivalent to the saturation field of the TLS, $\delta_0$ is the next dominant loss rate, and the filling factor $F$ is the ratio of electric field threading TLS to the total electric field.  $F\delta^0_{TLS}$ is the TLS loss tangent which is sensitive to a narrow spectrum of resonant TLS. Finally, $\beta$ describes the strength of TLS saturation with power. Early TLS theory suggests $\beta = 0.5$, however, recent results~\cite{burnett2014, burnett2017, kirsh2017, degraaf2017} commonly find a weaker scaling and associate it to a breakdown of the model described by Eq.~(\ref{eq:TLSpow}) due to interactions between TLS. Therefore, we allow $\beta$ to be a fit parameter initialized to $\beta = 0.5$. We find $\beta \simeq 0.2$ (see Table~\ref{tab:TLS}).

\begin{table}
    \centering
    \caption{\label{tab:TLS}Nanowire superinductance resonator parameters. $F\delta^0_{TLS}$ and $\beta$ are obtained from fits to Eq.~(\ref{eq:TLSpow}), $F\delta^i_{TLS}$ from fits to Eq.~(\ref{eq:pound}), and finally $P_\gamma$ from fits to Eq.~(\ref{eq:TLSlog}).}
    \begin{ruledtabular}
    \begin{tabular}{ccccc}
        NW $f_r$ & \multirow{2}{*}{$\beta$} & $F\delta^0_{TLS}$ & $F\delta^i_{TLS}$ & \multirow{2}{*}{$P_\gamma$} \\
        (\si{\mega\hertz}) & & ($\times 10^{-5}$) & ($\times 10^{-5}$) &  \\ \colrule
         4420 & 0.198 & 8.81 & 4.37 & 0.195 \\
         4562 & 0.196 & 7.51 & 3.84 & 0.183 \\
         4685 & 0.187 & 10.7 & 3.53 & 0.218 \\
         4837 & 0.180 & 8.24 & 4.40 & 0.153 \\
         5285 & 0.184 & 10.8 & 4.12 & 0.213
    \end{tabular}
    \end{ruledtabular}
\end{table}

\begin{figure}
    \includegraphics[width=8.5cm]{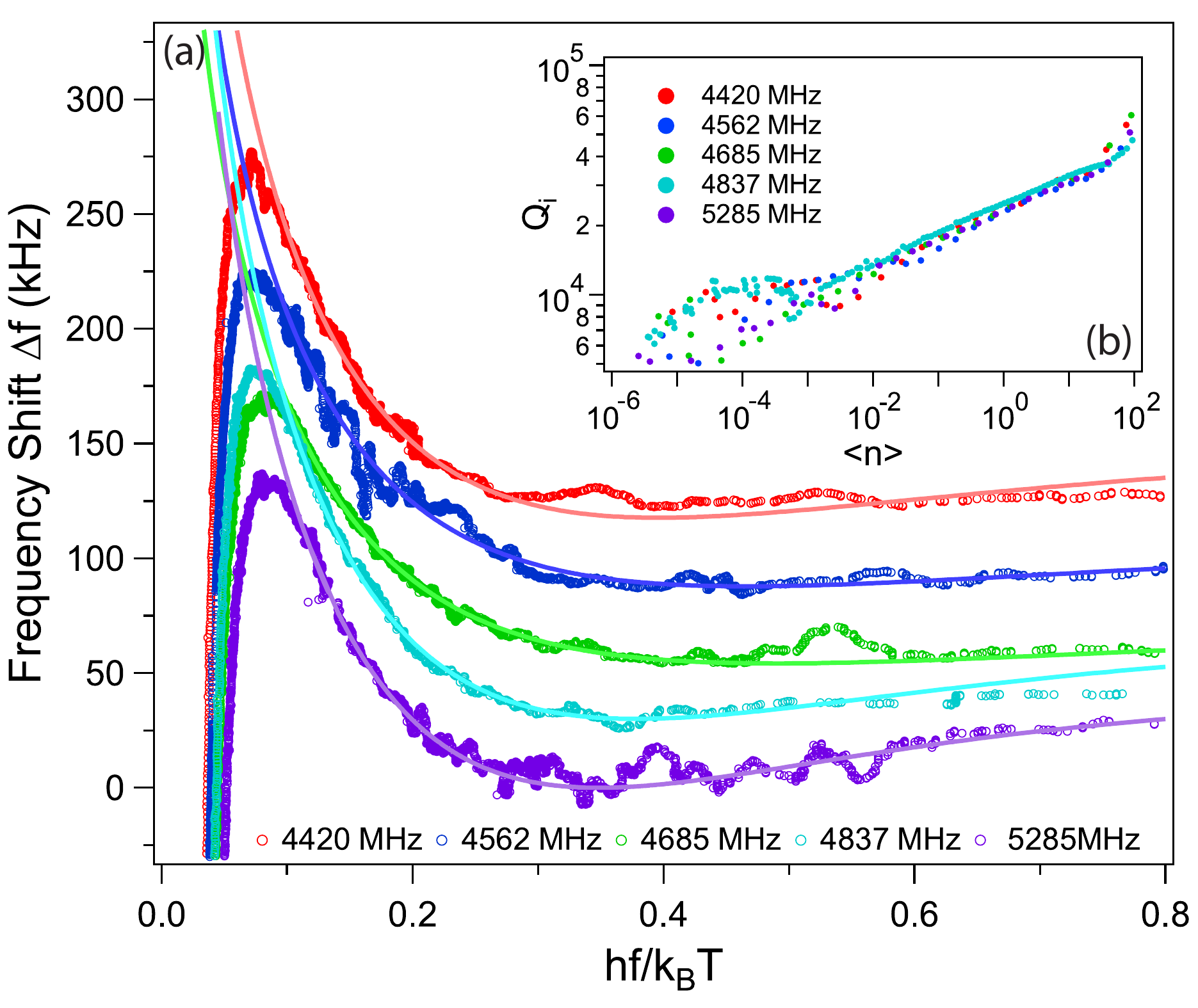}
    \caption{\label{fig:pound}{\bf (a)} Frequency shift as a function of the normalized frequency of all the nanowire superinductor resonators. The solid lines show fits to the theory described by Eq.~(\ref{eq:pound}). For clarity, the curves have been offset by $\SI{30}{\kilo\hertz}$. {\bf (b)} Shows $Q_i$ as a function of microwave drive for the same nanowire superinductor resonators (cf.\@ Fig.~\ref{fig:sweep}).}
\end{figure}

The intrinsic TLS loss tangent $F\delta^i_{TLS}$~\cite{gao2008}, which is sensitive to the complete TLS spectrum, can be unambiguously determined using a Pound frequency-locked loop~(P-FLL)\cite{burnett2014,degraaf2017}. The P-FLL  continuously tracks frequency changes of the resonator against temperature \footnote{see  Supplemental  Material  for  details  of  the  experimental  setup}. Figure~\ref{fig:pound}(a) shows the changes in resonance frequency against the natural energy scale of the TLS ($h f / k_B T$). The frequency shift~\cite{gao2008} is described by
\begin{equation}
\label{eq:pound}
\Delta f = F\delta^i_{TLS} \left( \ln\left( \dfrac{T}{T_0} \right) - \left[ g(T,f) - g(T_0,f) \right] \right)
\end{equation}
\noindent where $\Delta f = (f_r(T) - f_r(T_0))/f_r(T_0)$, $g(T,f) = \operatorname{Re}\left( \Psi\left( \frac{1}{2} + h f/2\pi i k_B T \right) \right)$, $T_0$ is a reference temperature and $\Psi$ is the complex digamma function. Importantly, Eq.~(\ref{eq:pound}) only fits the TLS contribution but does not fit the temperature-dependent kinetic inductance contribution which occurs below $h f / k_B T = 0.1$.

We use Eq.~(\ref{eq:pound}) to fit the measured data. The resulting values of $F\delta^i_{TLS}$ can be found in Table~\ref{tab:TLS}. We note that $F\delta^0_{TLS}$ and $F\delta^i_{TLS}$ differ by a factor 2 to 3, which is significantly larger than the $\sim 15\%$ difference typically observed~\cite{burnett2017}. However, this is not unexpected as the low value of $\beta$ indicated that we are outside of the range of validity of Eq.~(\ref{eq:TLSpow})\cite{burnett2017,degraaf2017}. The intrinsic loss tangent $F\delta^i_{TLS}$ can be used to fit the data in Figure~\ref{fig:sweep} using a model that takes TLS interactions into account~\cite{faoro2015,burnett2014}.
\begin{equation}
\label{eq:TLSlog}
\dfrac{1}{Q_i} = F \delta^i_{TLS} P_\gamma \ln\left( \dfrac{c n_c}{\left<n\right>} + \delta'_{0} \right) \tanh\left( \dfrac{h f_r}{2 k_B T} \right)
\end{equation}
\noindent where $c$ is a large constant, $\delta'_0$ is the log-scaled next dominant loss rate and $P_\gamma$ is the TLS switching rate ratio, defined by $P_\gamma = 1/\ln(\gamma_{max} / \gamma_{min})$ where $\gamma_{max}$ and $\gamma_{min}$ are the maximum and minimum rate of TLS switching respectively. These rates have been measured in the TLS-related charge-noise spectrum of single-electron transistors. They were found to extend from $\gamma_{min} \simeq \SI{100}{\hertz}$ to $\gamma_{max} \simeq \SI{25}{\kilo\hertz}$~\cite{kafanov2008}. This corresponds to $P_\gamma = 0.18$. Our fitted values of $P_\gamma$ are summarized in Table~\ref{tab:TLS}: we find values between $0.153$ and $0.218$, in good agreement with this estimate and other results~\cite{burnett2016, kirsh2017, degraaf2017}.

We have demonstrated that dissipation in our nanowires is not an intrinsic property of disorder within the film\cite{coumou2013microwave,feigel2017microwave} but is instead caused by TLS. This is not surprising as TLS are the predominant source of dissipation and decoherence in a wide variety of quantum devices. An important consequence of this is the role of the TLS filling factor. The ratio of E field threading TLS to the total E field is known to scale as approximately $1/\bar{w}$\cite{gao2008}, where $\bar{w}$ is the center conductor width of a superconducting resonator. Consequently, the 40~nm width used here to produce a superinductor leads to an unfavorable filling factor and therefore a much lower $Q_{\rm i}$ than is found for wider superconducting resonator geometries. Additionally, the nanowire lithography relies on the use of a spin-on glass resist (HSQ) which resembles amorphous silicon oxide. Silicon oxide is a well-known host of TLS~\cite{barends2008} and because some HSQ remains unetched atop our nanowires, we suspect this is the dominant source of TLS in our devices. Therefore, improvements to the fabrication, specifically the non-trivial removal of the HSQ mask should result in significant improvements in device performance. 

In conclusion, we have demonstrated a nanowire superinductance with an impedance of $\SI{6.795}{\kilo\ohm}$ and a single photon quality factor of $Q_i = \SI{2.5e4}{}$. This quality factor is comparable to both JJA-based superinductors\cite{masluk2012} and the tanh-scaled TLS loss in similar nanowire-resonators\cite{samkharadze2016}. We have analyzed the loss mechanisms in our devices and find TLS to be the dominant cause of loss, which is in contrast to the high rates of dissipation found in other nanowire\cite{astafiev2012,peltonen2013} or strongly disordered thin film devices~\cite{coumou2013microwave}. We emphasize that demonstrating nanowires losses are ``conventional'' is a important step forward for all nanowire-based quantum circuits. Therefore, this enables the possibility of long-lived nanowire-based superconducting circuits, such as a nanowire fluxonium qubit\cite{pop2014,kerman2010} or improved phase-slip qubits~\cite{astafiev2012,peltonen2013}. These high inductance and high impedance circuits can also be used as photon detectors~\cite{gao2008} or for exploring fundamental physics such as Bloch oscillations of charge for metrology applications~\cite{guichard2010}, or for increasing zero-point fluctuations for detection applications~\cite{samkharadze2016}.

\begin{acknowledgments}
The authors thank O. W. Kennedy for He FIB imaging of our devices. We acknowledge useful discussions with S.\@ E.\@ Kubatkin, A.\@ V.\@ Danilov, and P.\@ Delsing as well as support from the Chalmers Nanofabrication Laboratory staff. This research has been supported by funding from the Swedish Research Council and Chalmers Area of Advance Nanotechnology.
\end{acknowledgments}


%

%
%

\onecolumngrid
\clearpage
\begin{center}
\textbf{\large {Supplemental Material}}
\end{center}

\setcounter{equation}{0}
\setcounter{figure}{0}
\setcounter{table}{0}
\setcounter{page}{1}
\makeatletter
\renewcommand{\theequation}{S\arabic{equation}}
\renewcommand{\thefigure}{S\arabic{figure}}
\renewcommand{\thetable}{S\arabic{table}}
\renewcommand{\bibnumfmt}[1]{[S#1]}
\renewcommand{\citenumfont}[1]{S#1}

\section{Sample Fabrication}

Samples are fabricated on high-resistivity ($\rho \geq \SI{10}{\kilo\ohm\centi\meter}$) (100) intrinsic silicon substrates. Before processing, the substrate is dipped for $\SI{30}{\second}$ in hydrofluoric acid (HF) to remove any surface oxide. Within $\SI{5}{\minute}$, the wafer is loaded into a UHV sputtering chamber where a $\SI{20}{\nano\meter}$ thick NbN thin film is deposited by reactive DC magnetron sputtering from a $\SI{99.99}{\percent}$ pure Nb target in a 6:1 Ar:N$_{\text{2}}$ atmosphere. A $\SI{500}{nm}$-thick layer of PMMA A6 resist is spin-coated and then exposed with electron beam lithography (EBL) to define the microwave circuitry. After development, the pattern is transferred to the film by reactive ion etching (RIE) in a 50:4 Ar:Cl$_{\text{2}}$ plasma at $\SI{50}{\watt}$ and $\SI{10}{\milli\torr}$. The nanowires are patterned in a subsequent EBL exposure using a $\SI{50}{\nano\meter}$ layer of hydrogen silsesquioxane (HSQ), an ultra-high resolution negative resist suitable for $\leq \SI{10}{\nano\meter}$ features~\cite{chen2006}. \\ 

\begin{figure}[h]
    \centering
    \includegraphics[width=17cm]{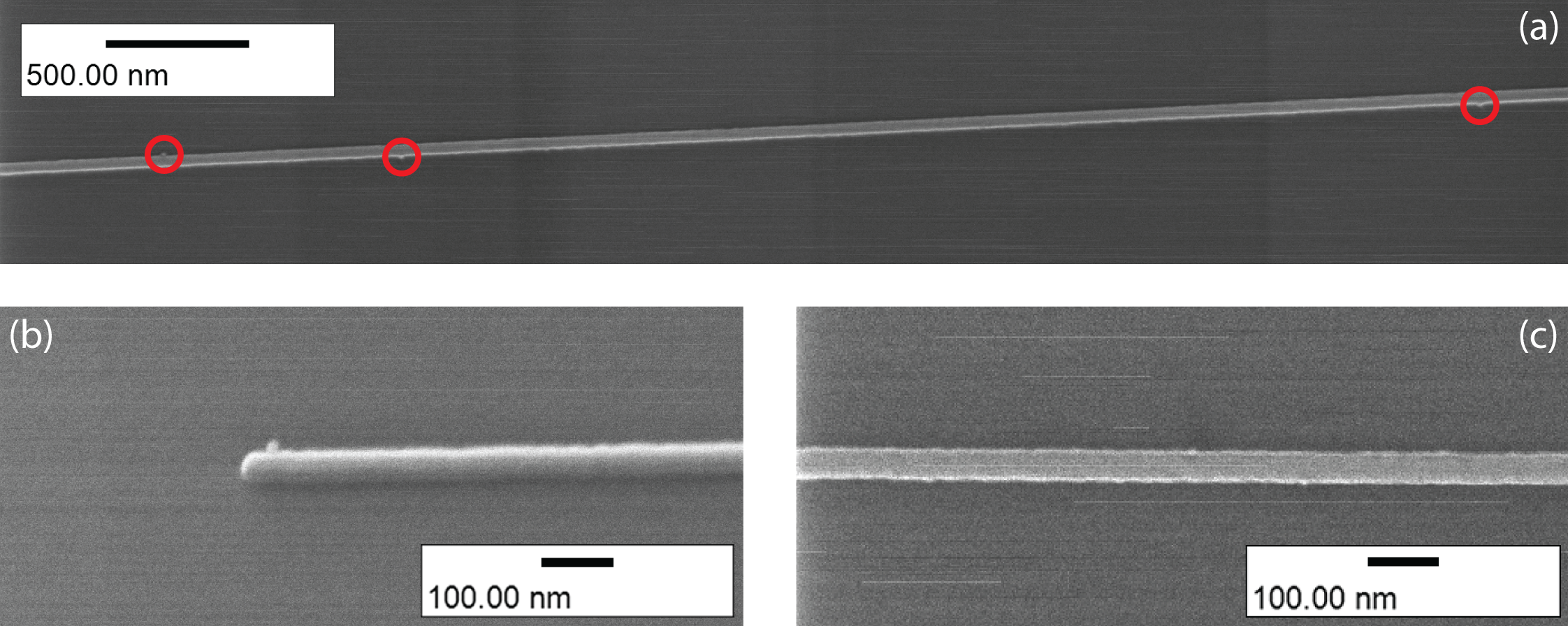}
    \caption{\label{fig:supp_fib} Helium FIB micrographs of a nanowire superinductor. \textbf{(a)} Shows a low magnification image of a long section ($\SI{5.5}{\micro\meter}$) of the nanowire. Lithographic defects are circled in red. \textbf{(b)} Shows a high magnification image of the end of the nanowire. Here we can see the defect is approximately $\SI{10}{\nano\meter}$ wide. \textbf{(c)} Shows a high magnification image of a section of nanowire without defects. Here we see the edge roughness is approximately $\pm \SI{1}{\nano\meter}$.}
\end{figure}

A common problem with HSQ is the formation of small agglomerates that are not completely dissolved during development. These small particles tend to accumulate on the edges of developed structures and act as micro-masks when the pattern is etched. From FIB micrographs of our devices (see figure~\ref{fig:supp_fib}), we estimate a lithographic defect rate less than 3 defects per $\SI{10}{\micro\meter}$ and that each defect contributes to the geometry of the device as approximately one square. For a $\SI{680}{\micro\meter}$ long and $\SI{40}{\nano\meter}$ nanowire, this translates to an uncertainty upper bound of $\SI{17000}{\sq} \pm \SI{200}{\sq}$, which is consistent with the 1\% error reported in the main text.

\section{Exponentially suppressed phase slip rate}

In the main text, we make the argument that the device dimensions are chosen to exponentially suppress phase slips. We estimate the phase slip rate $\Gamma_S = E_S / h = E_0 / h \exp(-\kappa \bar{w})$ for our device within the phenomenological model for strongly disordered superconductors. Our analysis is similar to that of Peltonen \textit{et al.}~\cite{peltonen2013sup}. In this model, $E_S$ is the phase slip energy and we have $E_0 = \rho \sqrt{l/\bar{w}}$, where $l$ and $\bar{w}$ are the nanowire length and average width respectively, $\rho = (\hbar/2e)^2 / L_k^{\Box}$ represents the superfluid stiffness, $\kappa = \eta \sqrt{\nu_p \rho}$, $\eta \simeq 1$ and $\nu_p = 1/(2e^2 R_N^{\Box} D)$ is the Cooper pair density of states with $D \simeq \SI{0.45}{\centi\meter\squared\per\second}$.\\

For our device parameters (summarized in table~\ref{tab:params}), we find $\Gamma_S \simeq \SI{7e-5}{\hertz}$. 

\begin{table}[h]
    \centering
    \caption{\label{tab:params}Nanowire superinductance film and device parameters.}
    \begin{tabular}{ccc}
        \hline\hline
        Parameter & Symbol & Value  \\ \colrule
        Normal state sheet resistance & $R_N^{\Box}$ & $\SI[per-mode=symbol]{503}{\ohm\per\sq}$ \\
        Sheet kinetic inductance & $L_k^{\Box}$ & $\SI[per-mode=symbol]{82}{\pico\henry\per\sq}$ \\
        Nanowire length & $l$ & $\SI{680}{\micro\meter}$ \\
        Nanowire width & $\bar{w}$ & $\SI{40}{\nano\meter}$ \\
        \hline\hline
    \end{tabular}
\end{table}

\section{Measurement Setup}
\begin{figure}[h]
    \includegraphics[width=18cm]{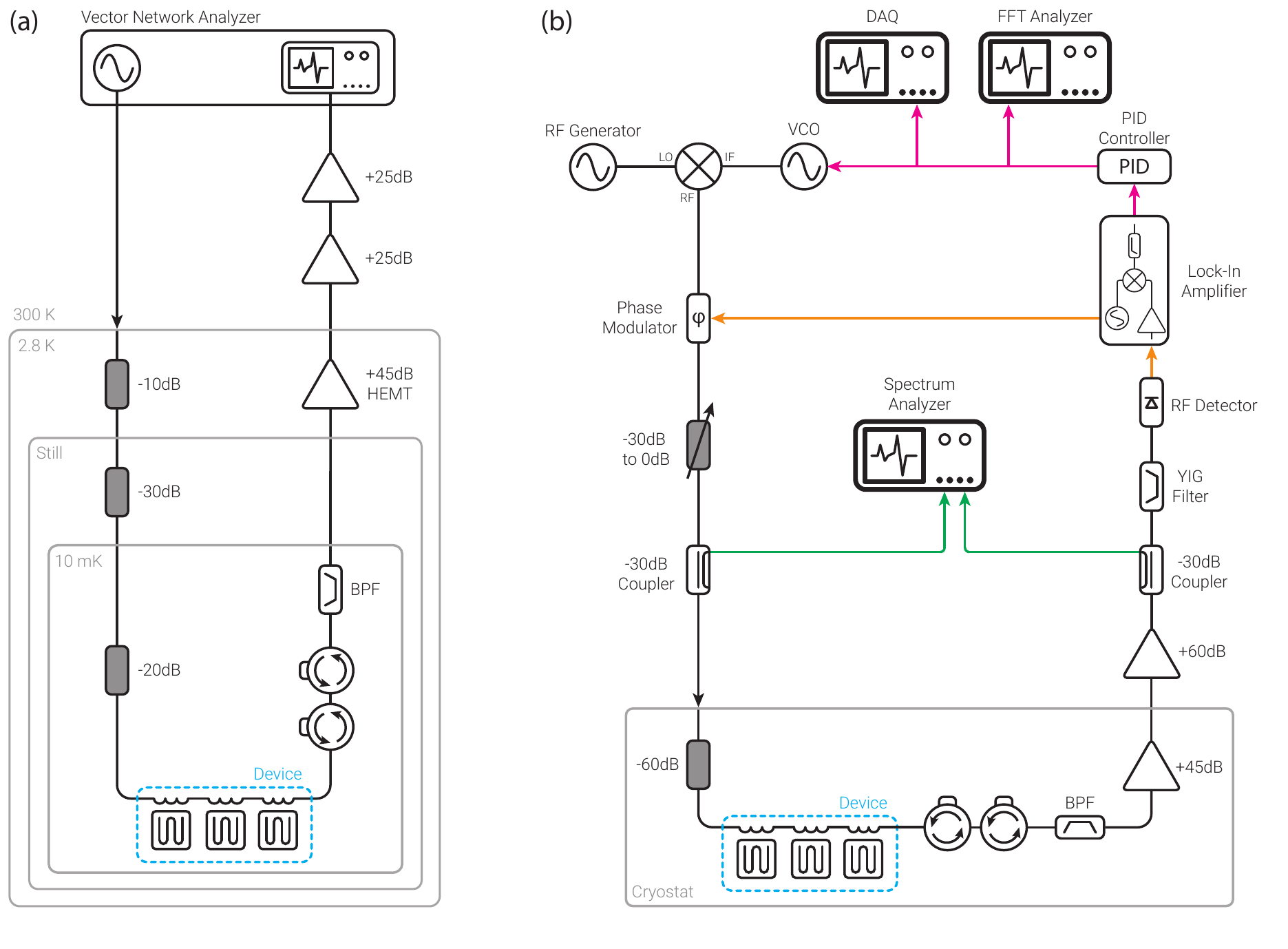}
    \caption{\label{fig:sup_setup}(a) Cryogenic microwave setup. (b) Schematic of the Pound frequency locked loop (P-FLL).}
\end{figure}

The sample is wire-bonded in a connectorized copper sample-box that is mounted onto the mixing chamber stage of a Bluefors LD250 dilution refrigerator (Fig.~\ref{fig:sup_setup}a). The inbound microwave signal is attenuated at each temperature stage by a total of $\SI{60}{\decibel}$ before reaching the device under test. Accounting for cable losses and sample-box insertion loss, the total attenuation of the signal reaching the sample is $\SI{70}{\decibel}$. To avoid any parasitic reflections and noise leakage from amplifiers, the transmitted signal is fed through two microwave circulators (Raditek RADI-4.0-8.0-Cryo-4-77K-1WR) and a 4-8 \si{\giga\hertz} band pass filter. Finally, the signal is amplified by a LNF LNC4\_8A HEMT cryogenic amplifier ($\SI{45}{\decibel}$ gain) installed on the $\SI{2.8}{\kelvin}$ stage. Additional amplification is done at room temperature (Pasternack PE-1522 gain block amplifiers).\\

This microwave setup is connected to a vector network analyzer (Keysight PNA-X N5249A or R\&S ZNB20) for initial characterization and quality factor measurements of the nanowire resonators at various excitation powers (Fig.~\ref{fig:sup_setup}a). However, as highlighted in the main text, at low drive powers, VNA measurements require significant amounts of averaging to increase the signal-to-noise ratio (SNR). At low microwave energies, frequency jitter leads to spectral broadening.\\


To reliably determine the TLS loss contribution, we instead measure the resonance frequency of the resonator against temperature~\cite{gao2008sup, lindstrom2009}. For that purpose, the microwave setup is included in a frequency locked loop using the so-called Pound locking technique (Fig.~\ref{fig:sup_setup}b). Originally developed for microwave oscillators~\cite{pound1946}, this technique is commonly used in optics for frequency stabilization of lasers~\cite{black2001} and has been recently used for noise~\cite{lindstrom2009, lindstrom2011} and ESR~\cite{degraaf2012, degraaf2017sup} measurements with superconducting microwave resonators. In this method, a carrier signal is generated by mixing the output of a microwave source (Keysight E8257D) and a VCO (Keysight 33622A). This carrier is phase-modulated (Analog Devices HMC538) before being passed through the resonator under test. The phase modulation frequency is set so that the sidebands are not interacting with the resonator. After amplification, the signal is filtered (MicroLambda MLBFP-64008) to remove the unwanted mixer image and rectified using an RF detector diode (Pasternack PE8016). The diode output is demodulated with a lock-in amplifier (Zurich Instruments HF2LI).\\

The feedback loop consist of an analog PID controller (SRS SIM960) locked on the zero-crossing of the error signal. This gives an output directly proportional to any shift in resonance frequency of the resonator. This output signal is then used to drive the frequency modulation of the VCO, varying its frequency accordingly and enabling the loop to be locked on the resonator.\\

In this work, we only sample the PID output slowly ($\leq \SI{100}{\hertz}$) to track frequency changes (Keithley 2000), but noise in the resonator can also be studied using a frequency counter (Keysight 53132A), a fast-sampling DAQ (NI PXI-6259 DAQ) or an FFT analyser (Keysight 35670A). This will be the focus of future work.

\section{Geometrical considerations for nanowire superinductors}

\begin{figure*}[h]
    \includegraphics[width=11cm]{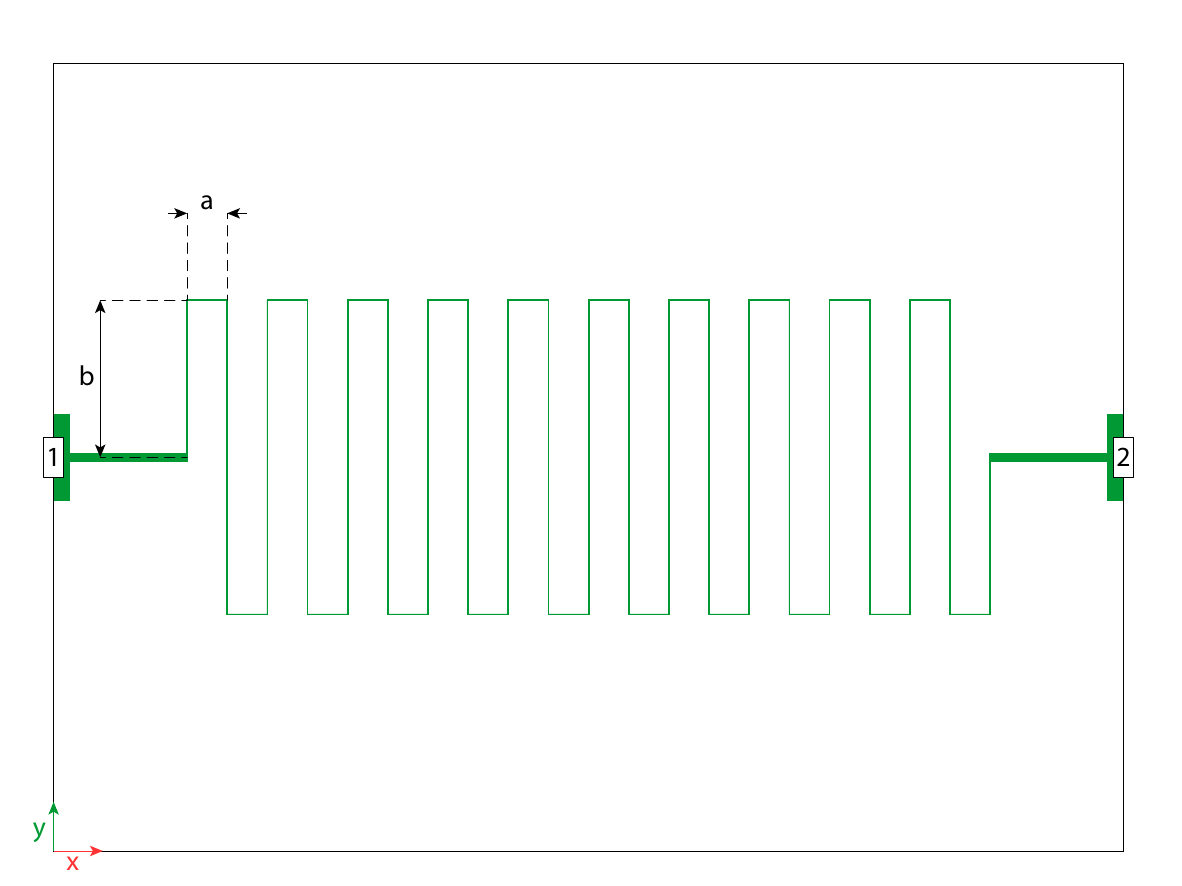}
    
    \caption{\label{fig:sup_struct} Schematic representation of a typical meandered nanowire resonator structure simulated in this section. $a$ and $b$ are the distance between meanders and the meander length respectively. $1$ and $2$ are the excitation and measurement ports and the black outline represents the grounded edge of the simulation box.}
\end{figure*}

In this section, we analyze the influence of meandering the nanowire to qualitatively study the role of any geometry dependent parasitic capacitance. For that purpose, we simulate the frequency response and current density of various nanowire superinductors using Sonnet \textit{em} microwave simulator. In order to reduce meshing and simulation times, we simulate $\SI{100}{\nano\meter}$-wide nanowires in a simple step-impedance resonator geometry. We start by simulating a straight nanowire as a reference and then proceed to simulate nanowires in a meandered geometry with a fixed meander length $b = \SI{20}{\micro\meter}$ while gradually decreasing distance between meanders from $a = \SI{30}{\micro\meter}$ (typical distance in our devices) to $\SI{100}{\nano\meter}$ (see figure~\ref{fig:sup_struct}).

\begin{figure*}[h]
    \includegraphics[width=11cm]{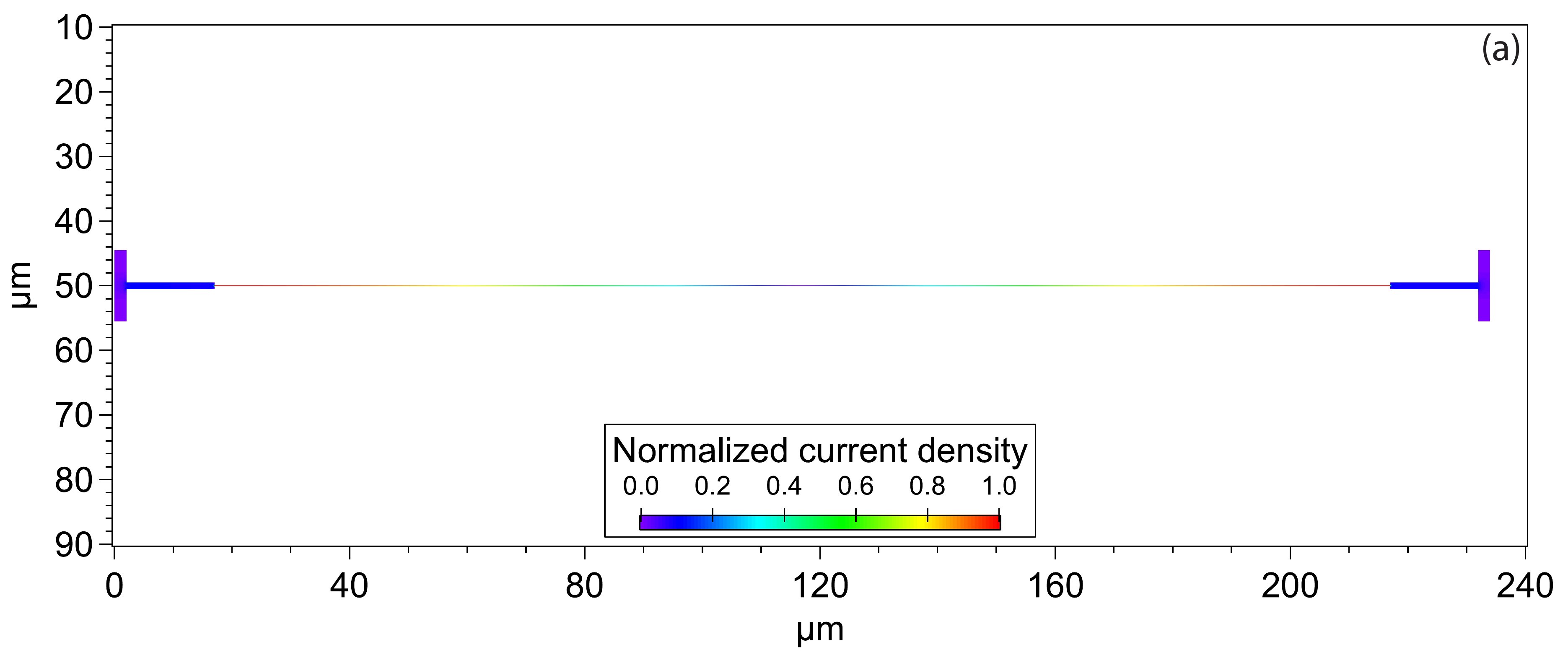}
    \includegraphics[width=5.5cm]{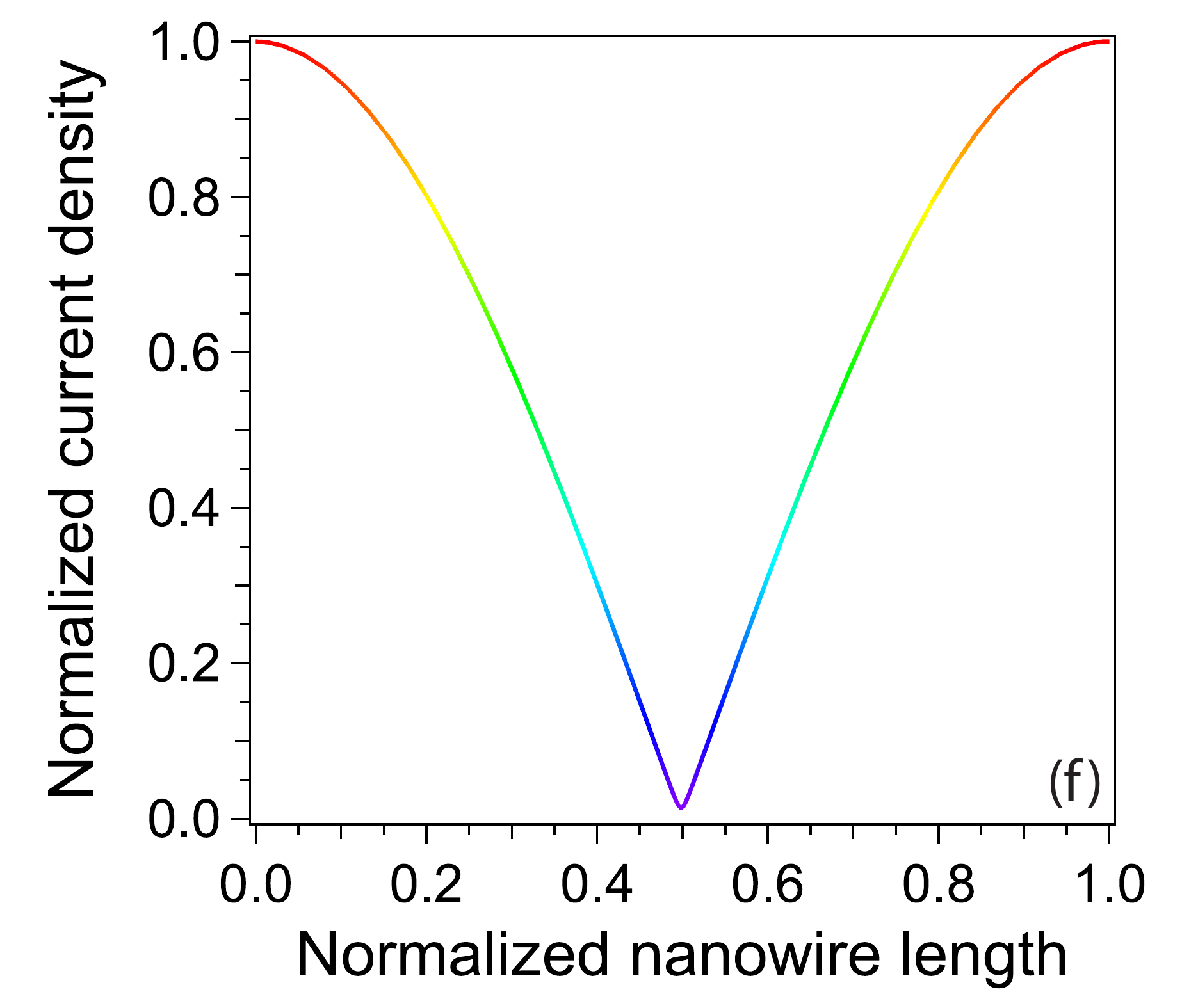}
    
    \includegraphics[width=11cm]{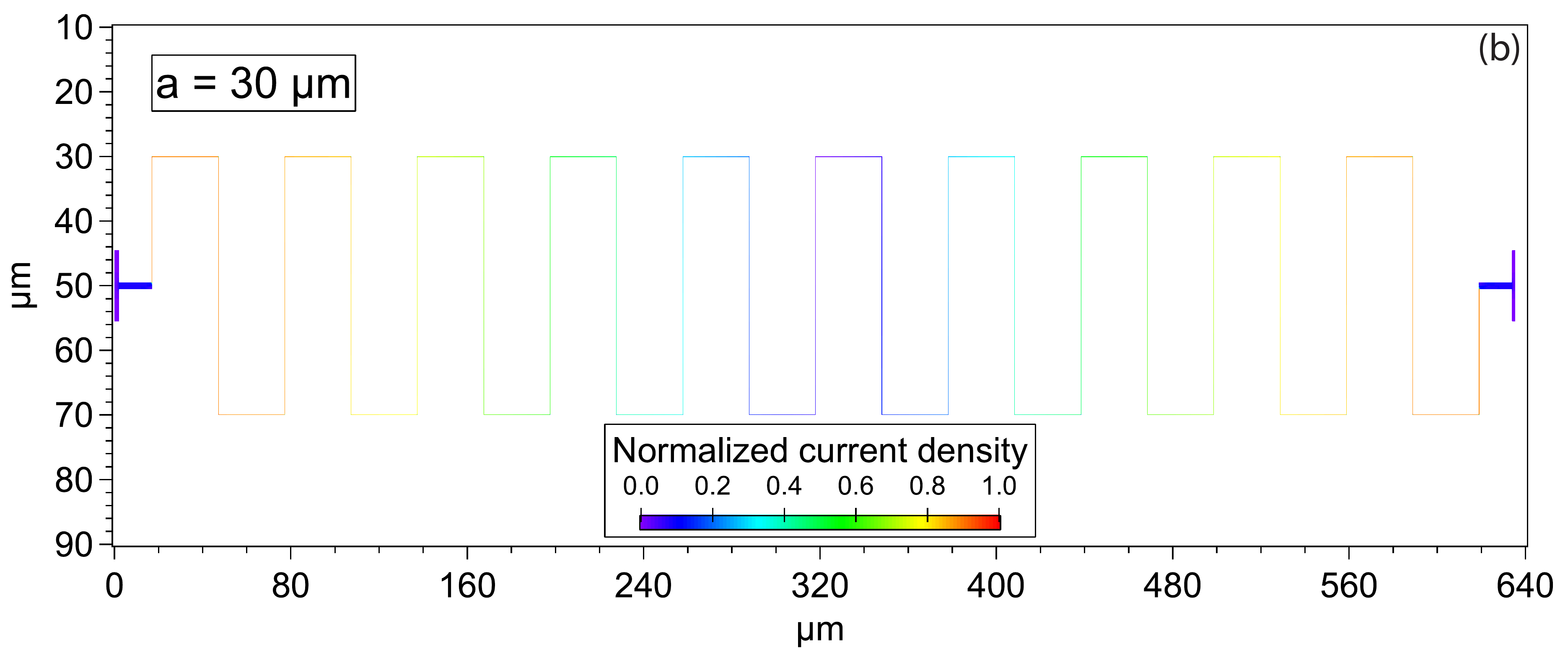}
    \includegraphics[width=5.5cm]{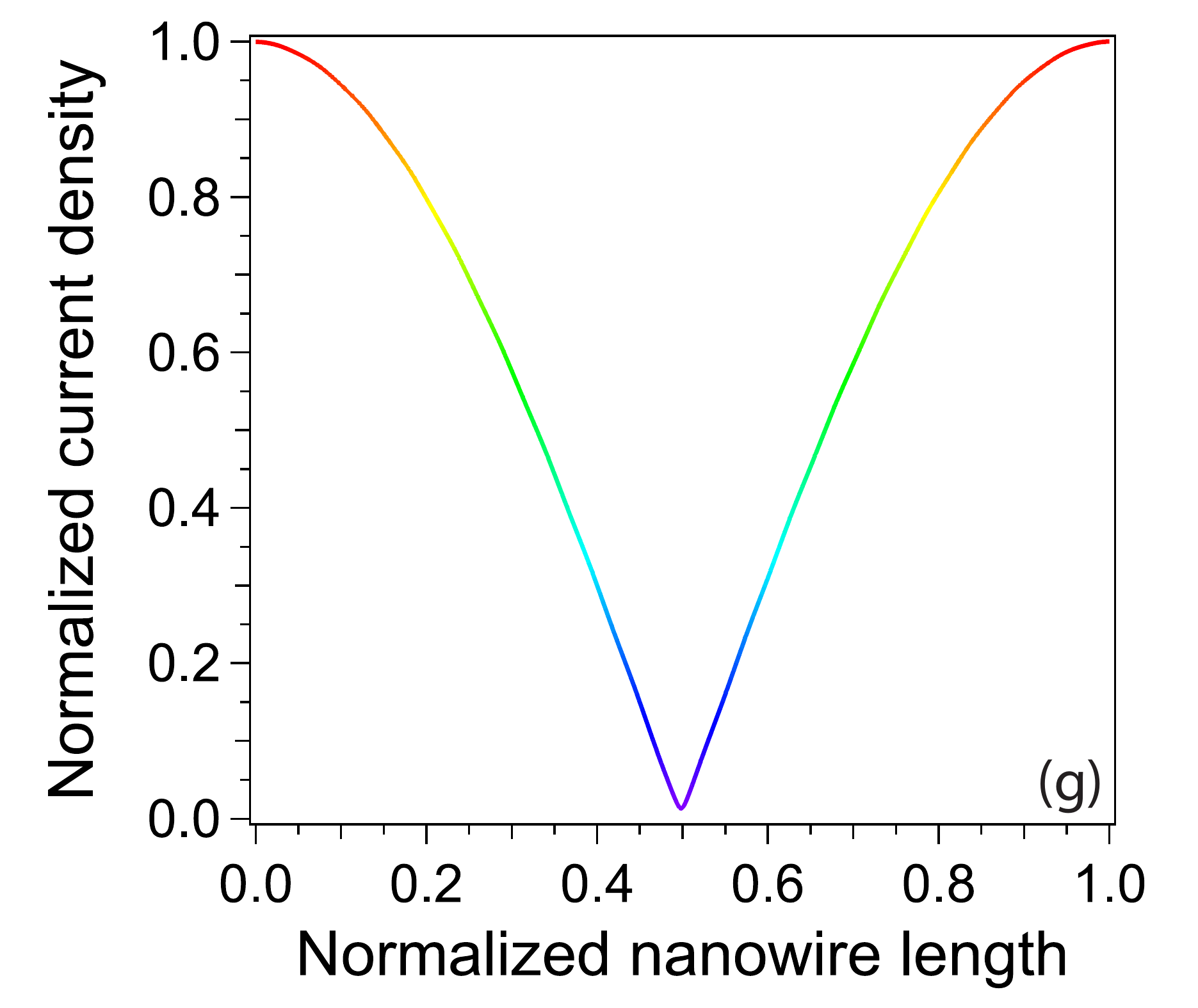}
    
    \includegraphics[width=11cm]{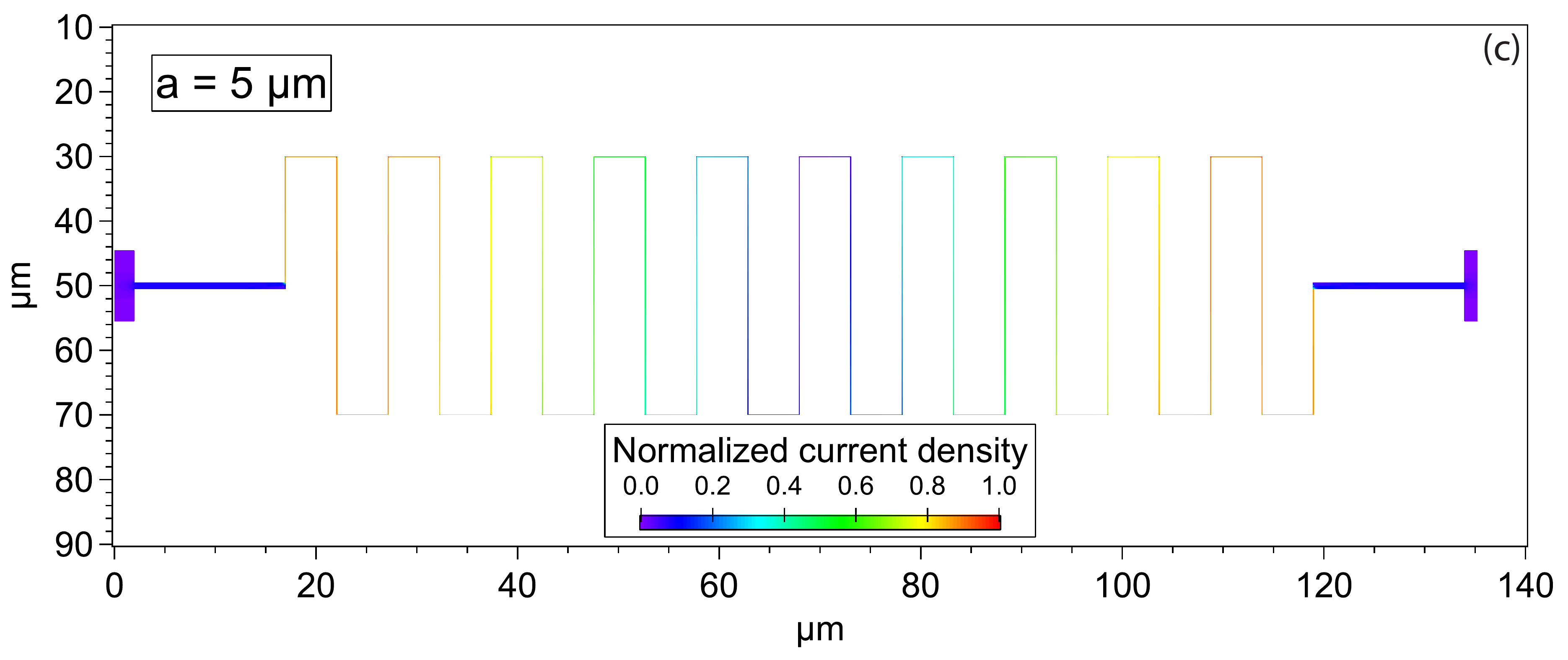}
    \includegraphics[width=5.5cm]{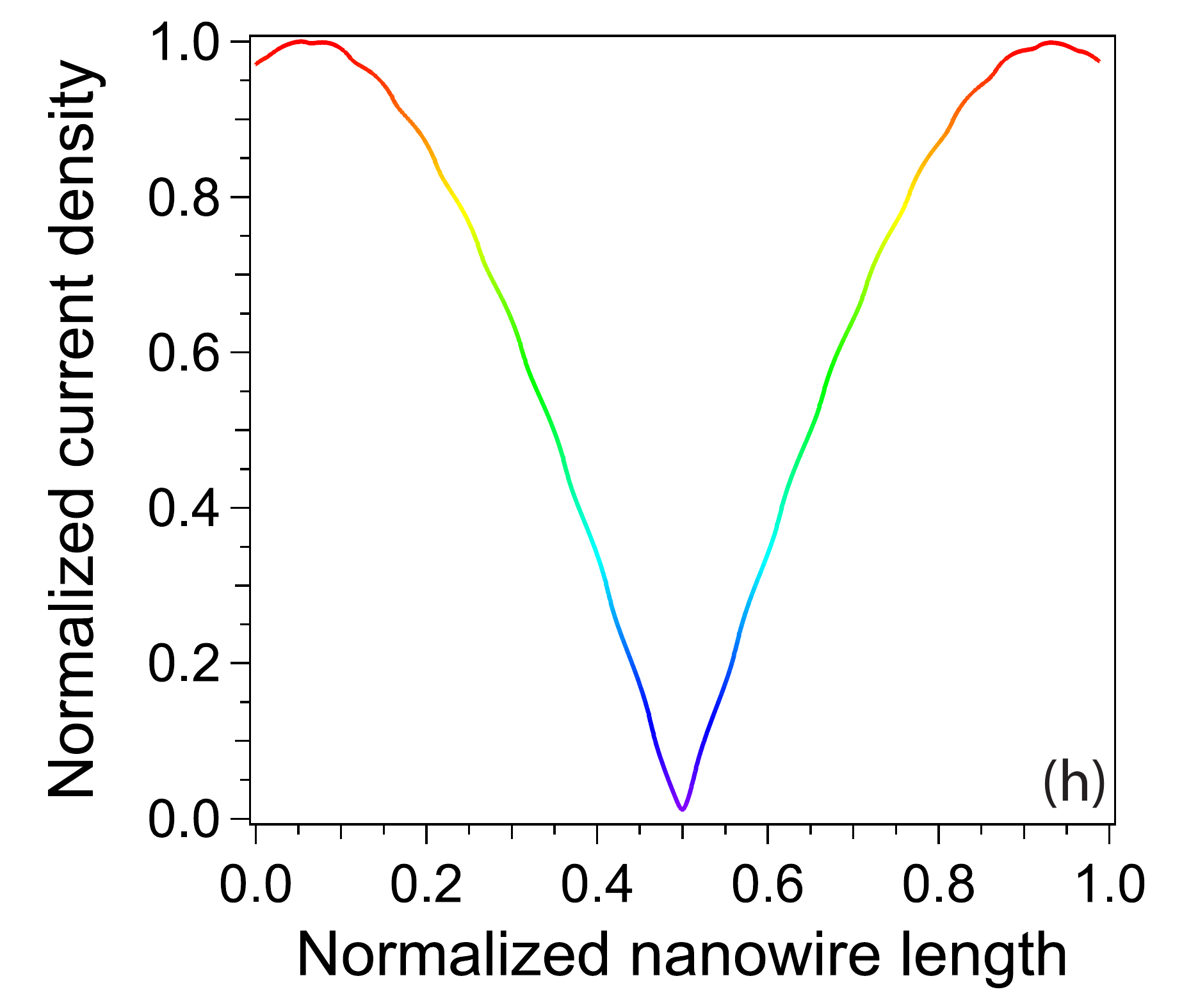}
    
    \includegraphics[width=11cm]{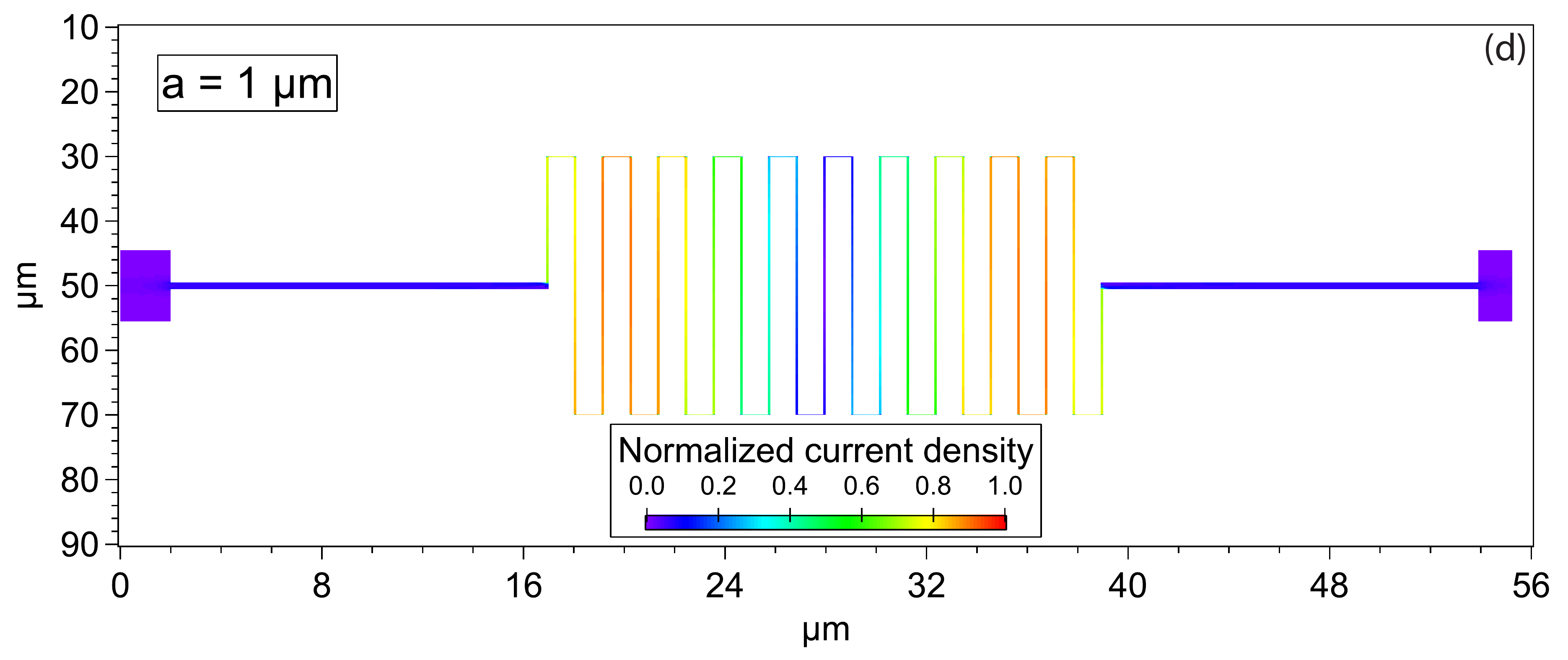}
    \includegraphics[width=5.5cm]{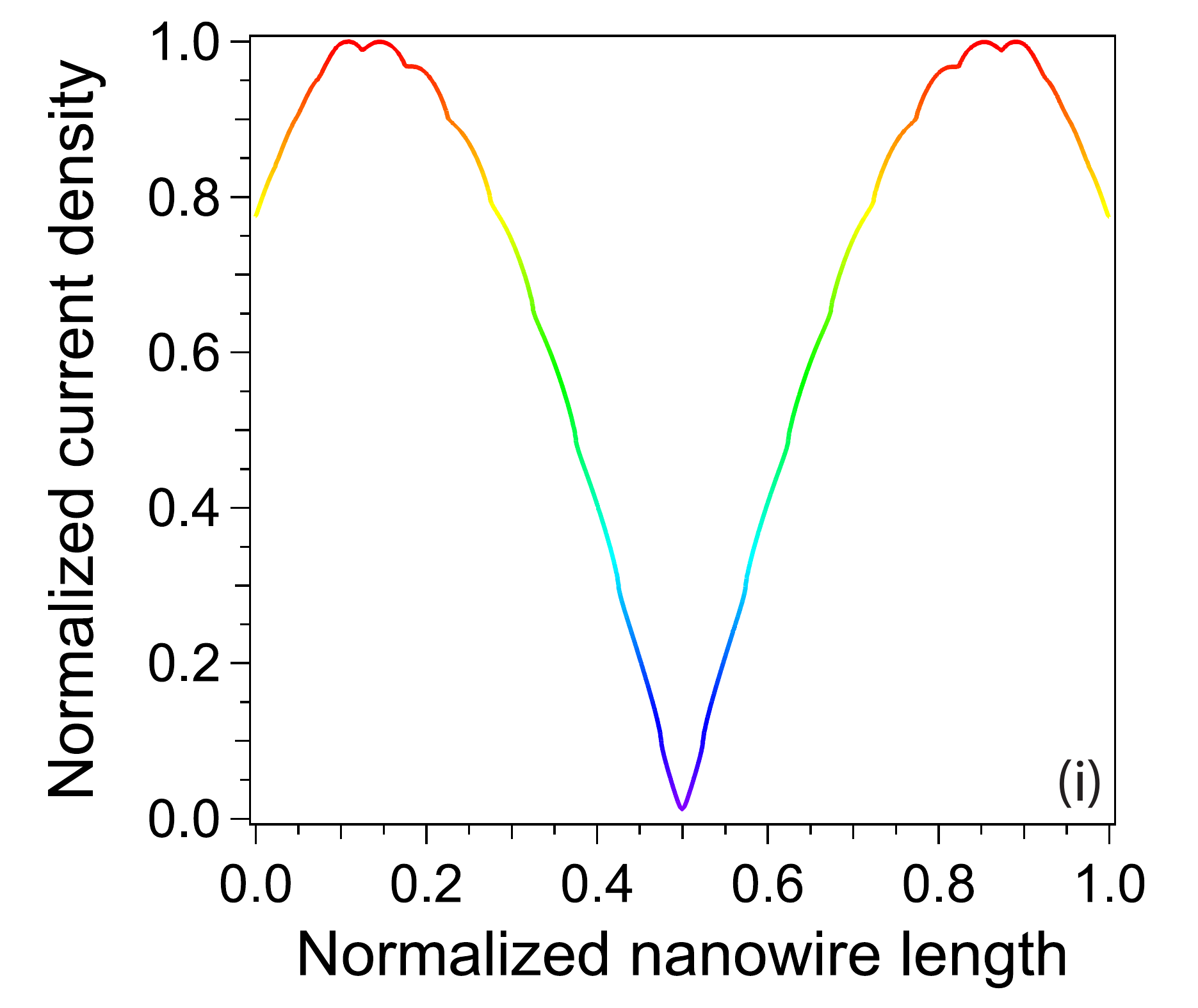}
    
    \includegraphics[width=11cm]{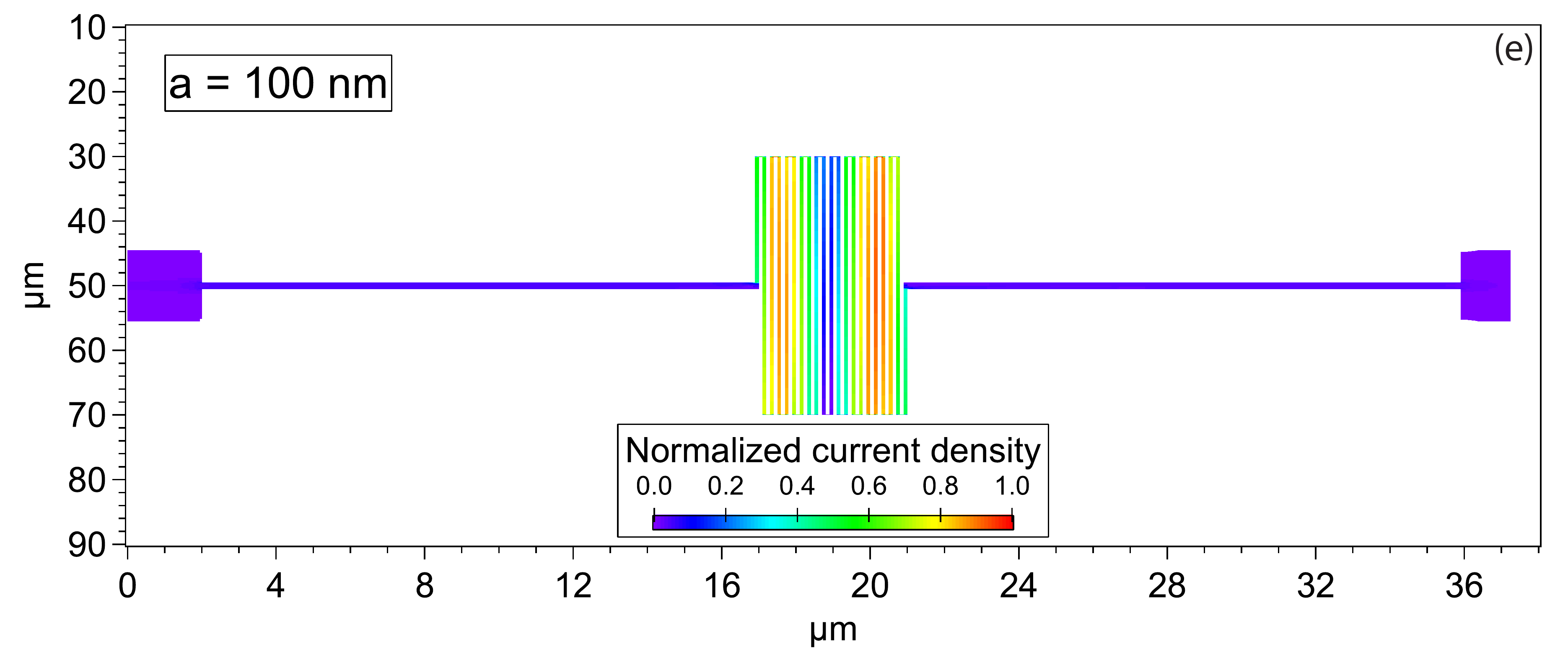}
    \includegraphics[width=5.5cm]{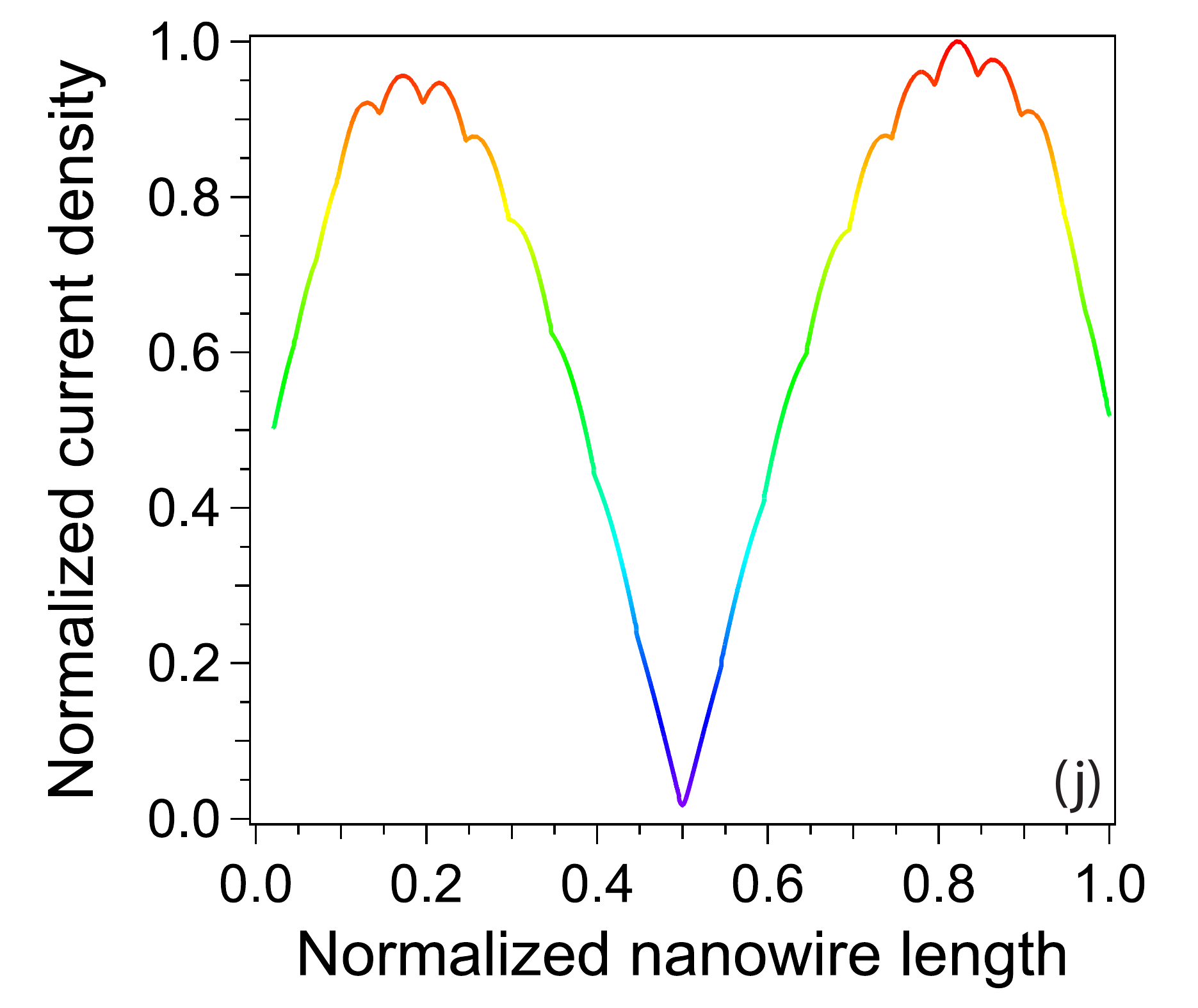}
    
    \caption{\label{fig:sup_density} \textbf{(a-e)} Current density distribution in nanowires at several inter-meander distances. \textbf{(f-j)} Corresponding normalized current density along the nanowire.}
\end{figure*}

\begin{figure*}[h]
    \includegraphics[width=10.75cm]{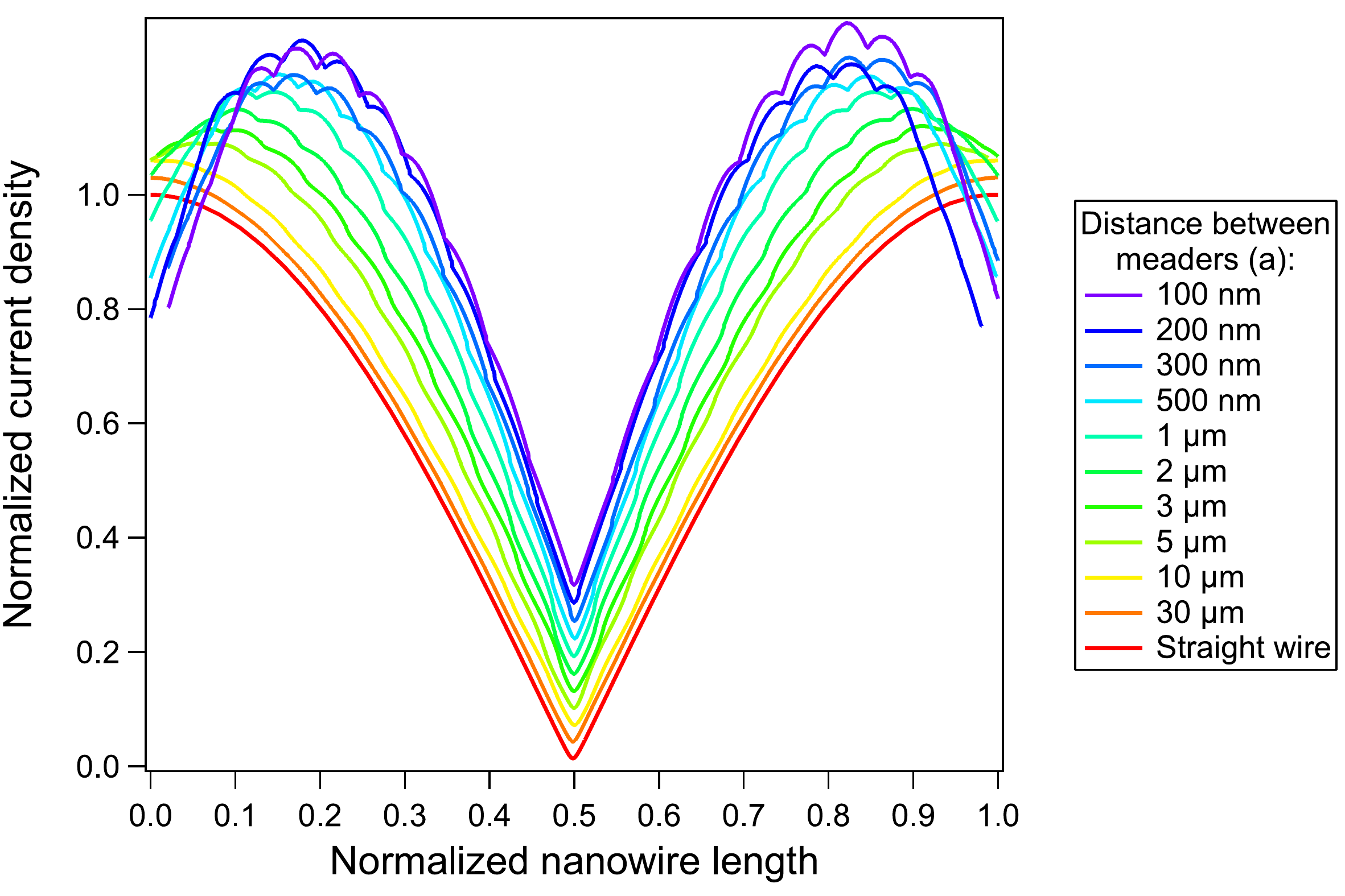}
    
    \caption{\label{fig:sup_all} Normalized current densities at the fundamental resonance frequencies for all the simulated structures. For clarity, the curves have been offset by 0.03.}
\end{figure*}

Figures~\ref{fig:sup_density} and~\ref{fig:sup_all} show the normalized current density along the nanowires at the fundamental resonance frequency of the simulated structure. To be clear, for meandered geometries, the current density is not measured as a line cut along the x-axis, but instead the geometry is unwound and the current density is extracted at every point along the nanowire. We observe that for the straight wire and for $a > \SI{5}{\micro\meter}$, the current density is consistent with the expected $\lambda/2$ mode structure of such a resonator and the characteristic impedance of the nanowire superinductor is well-defined to $Z = \sqrt{L_{nw}/C_{nw}}$ as described in the main text. \\

However, below $a = \SI{5}{\micro\meter}$, we observe that, as the distance reduces between the meanders, the resonance frequency significantly diverges from the straight nanowire reference value and the current density is severely distorted. This is explained by the increasing influence of parasitic capacitance between each meander. This parasitic capacitance is equivalent to shunting the nanowire with an extra capacitance and lowering its impedance. Moreover, the structure cannot be treated as a $\lambda/2$ resonator anymore and has therefore no well-defined wave impedance.

\clearpage

%

\end{document}